\documentclass[aps,prl,preprint,superscriptaddress]{revtex4-2}

\usepackage{graphicx}
\usepackage{siunitx}
\usepackage{amsmath}
\usepackage{float}
\usepackage[section]{placeins}

\begin{document}


\title{Correlated double-electron additions at the \\edge of a two-dimensional electronic system}


\author{Ahmet Demir}
\affiliation{Department of Physics, Massachusetts Institute of Technology, Cambridge, Massachusetts 02139, USA}

\author{Neal Staley}
\affiliation{Department of Physics, Massachusetts Institute of Technology, Cambridge, Massachusetts 02139, USA}

\author{Samuel Aronson}
\affiliation{Department of Physics, Massachusetts Institute of Technology, Cambridge, Massachusetts 02139, USA}

\author{Spencer Tomarken}
\affiliation{Department of Physics, Massachusetts Institute of Technology, Cambridge, Massachusetts 02139, USA}

\author{Ken West}
\affiliation{Department of Electrical Engineering, Princeton University, Princeton, New Jersey 08544, USA}

\author{Kirk Baldwin}
\affiliation{Department of Electrical Engineering, Princeton University, Princeton, New Jersey 08544, USA}

\author{Loren Pfeiffer}
\affiliation{Department of Electrical Engineering, Princeton University, Princeton, New Jersey 08544, USA}

\author{Raymond Ashoori}
\email{ashoori@mit.edu}
\affiliation{Department of Physics, Massachusetts Institute of Technology, Cambridge, Massachusetts 02139, USA}


\date{\today}

\begin{abstract}
    We create laterally large and very low disorder GaAs quantum well based quantum dots to study single electron additions to two dimensional electron systems (2DES). Electrons tunnel into these dots across an AlGaAs tunnel barrier from a single $n+$ electrode. Using single-electron capacitance spectroscopy in a dilution refrigerator, we identify capacitance peaks for the addition of the first electron to a dot and record subsequent peaks in the addition spectrum up to occupancies of thousands of electrons. Here, we report two remarkable phenomena that occur in the Landau level filling factor range $\nu=2$ to $\nu=5$ in selective probing of electron additions to the edge states of the dot: (1) Coulomb blockade peaks arise from the entrance of two electrons rather than one; (2) at and near filling factor 5/2 and at fixed gate voltage, these double-height peaks appear uniformly with a periodicity of $h/2e$. At other filling factors in the range $\nu=2$ to $\nu=5$, the mean periodicity for the twice-height electron peaks remains $h/2e$, but the twice-height peaks instead further bunch into pairs of double-height peaks. The unusual two-electron Coulomb blockade peaks suggest a novel pair tunneling effect that involves electron correlations arising in the quantum dot, with spectra at $\nu=5/2$  identical to those previously only seen in superconducting dots.
\end{abstract}


\maketitle

Coulomb repulsion dictates an increase in the amount of energy required to add each successive electron to an isolated quantum dot, resulting in distinct electron additions in a periodic pattern known as a Coulomb blockade spectrum\cite{glazman_coulomb_1989, beenakker_theory_1991, silsbee_comment_1990, ashoori_n-electron_1993, devoret_introduction_1992, ashoori_electrons_1996}. Such a uniform pattern is commonly seen in two-dimensional (2D) semiconductor quantum dots\cite{ashoori_electrons_1996, kastner_artificial_1993}. In contrast, in superconducting dots, Cooper pairing leads to deviations from ordinary Coulomb blockade, producing individual peaks in the addition spectrum that arise from the addition of two electrons\cite{lafarge_two-electron_1993, tuominen_experimental_1992, joyez_observation_1994}. The ground state of a superconducting island favors even numbers of electrons, resulting in parity-induced suppression of Coulomb blockade\cite{matveev_parity-induced_1993,hekking_coulomb_1993} and $2e$ tunneling.  Here, we present results from an experiment that, surprisingly, reveals $2e$ tunneling into edge states of large 2D quantum dots. Remarkably, only around Landau level filling factor 5/2, the electron additions to the edge states behave identically to the observed behavior for Cooper pairs entering a superconducting dot, with all electrons entering as pairs, with no additional energy cost for adding a second electron after the first. Moreover, the paired additions exist over a wide range of filling factors (from $\nu=2$ to $\nu=5$), but at filling factors other than 5/2, the double-additions themselves group into pairs of double-additions. This exact doubling of capacitance peaks means that the experiments cannot be described by means of uncorrelated tunneling of electrons. These unusual effects in semiconductor dots all suggest the existence of composite particles within the 2D system comprised of two electrons which, at filling factor 5/2, behave in these experiments identically to expectations for Cooper pairs\cite{lafarge_two-electron_1993}.

\begin{figure}
    \centering
    \includegraphics[width=0.7\textwidth]{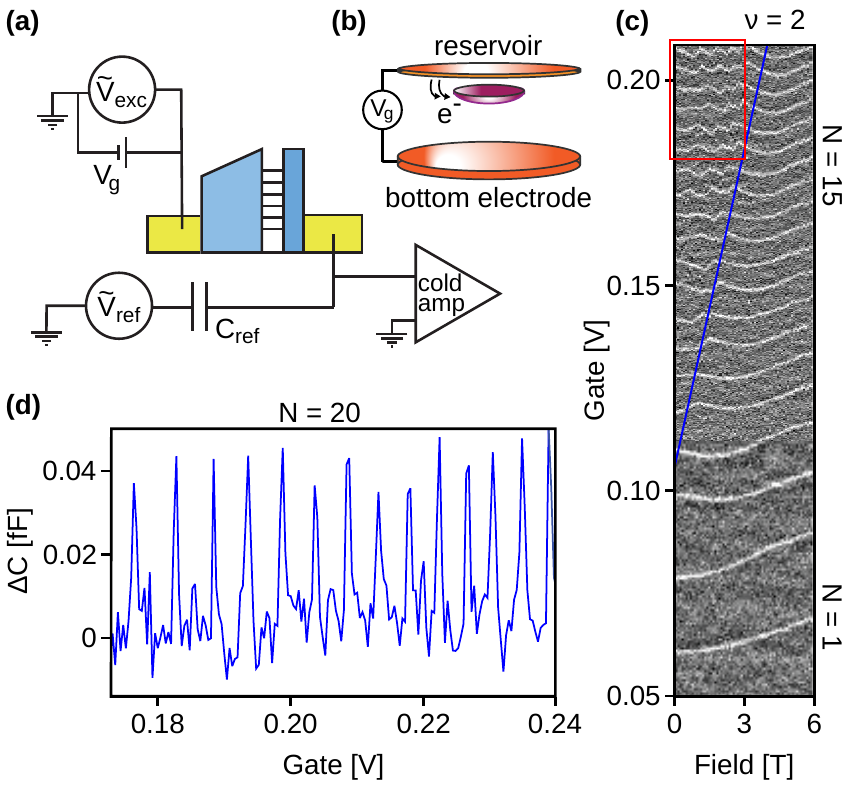}
    \caption{Electron additions in a small 2D quantum dot. (\textbf{a}) The quantum dot is confined between two electrodes, one of which (tunneling electrode) is tunnel coupled to the dot. The carrier concentration can be tuned by the gate electrode. Using a capacitance bridge, we balance the sample capacitor against a known reference capacitor. (\textbf{b}) Schematic of the sample design with a small metallic island (dot) that is tunnel coupled to a reservoir and capacitively coupled to the bottom electrode. (\textbf{c}) Single electron capacitance peaks in a small quantum dot as a function of external magnetic field and gate voltage. For electron numbers $N>15$, ``zig-zags'' appear in the traces arising from the lowest energy available electronic state shifting between the interior and the edge of the dot. (\textbf{d}) Isolated electron additions as a function of gate voltage. The spacing between successive capacitance peaks largely reflects the additional energy required to overcome the Coulomb repulsion of the electrons already in the dot.}
    \label{fig:capacitanceGeometry}
\end{figure}

Laterally confined two-dimensional (2D) quantum dots provide a simple system to study confined electrons and their interactions\cite{ashoori_electrons_1996,kouwenhoven_quantum_1998,oosterkamp_maximum-density_1999,tarucha_shell_1996}. However, conventional transport measurements on lateral quantum dots function by passing a current through them and can thereby only detect delocalized electronic states\cite{kastner_artificial_1993}. In order to access the single electron addition spectrum of laterally large quantum-confined structures, we utilize a vertical tunneling geometry that eliminates lateral contacts and allows us to track electron additions from the first up to thousands of electrons over a wide range of magnetic field\cite{ashoori_n-electron_1993,ashoori_single-electron_1992}. We have also developed a sample design to create quantum well-based large QDs with very low disorder. The dot is confined between two electrodes in the ``tunnel capacitor'' structure shown schematically in Figs.\ \ref{fig:capacitanceGeometry}a and \ref{fig:capacitanceGeometry}b. Unlike previous work, the structure does not contain any modulation doping\cite{dingle_electron_1978} nor a Schottky barrier above the dot. Instead, we have developed a new sample design that incorporates a small ohmic contact to a tunneling electrode incorporated in an etched pillar to eliminate the need for any unscreened dopants and populate the QD with electrons entirely by gating.

We detect electron additions as peaks in the capacitance of this tunnel capacitor using a capacitance bridge technique in which we balance the QD against a known reference capacitor\cite{ashoori_single-electron_1992}.  When a single electron tunnels from the tunnel electrode to the quantum dot, image charge accumulates on the opposite electrode. Using low temperature amplification in a dilution refrigerator, we read out the voltage created by this image charge to find the capacitance change\cite{ashoori_single-electron_1992}. The gate voltages for the observed capacitance directly reflect the ground state energies of the quantum dot containing different numbers of electrons\cite{ashoori_n-electron_1993,ashoori_electrons_1996,ashoori_single-electron_1992}. Previously, these and other electron addition spectroscopy measurements of quantum dots showed well-defined transitions into the integer quantum Hall states, an indication of electron interactions\cite{oosterkamp_maximum-density_1999,tarucha_shell_1996,zhitenev_localization-delocalization_1999,klein_exchange_1995}.

To illustrate the functioning of the capacitance method, we first show data in Figs.\ \ref{fig:capacitanceGeometry}c and \ref{fig:capacitanceGeometry}d from a small dot ($\sim$ $\SI{120}{\nano\meter}$) that does not show the pairing effect and follows standard ``artificial atom'' physics\cite{ashoori_electrons_1996,kouwenhoven_quantum_1998}. Electron additions occur periodically in gate voltage with a period of roughly $e/C_g$, where $C_g$ is the capacitance between the gate and the dot. After each electron is added to the dot, adding a successive electron requires increased gate voltage due to Coulomb repulsion\cite{beenakker_theory_1991,silsbee_comment_1990,ashoori_n-electron_1993,devoret_introduction_1992}. Fig.\ \ref{fig:capacitanceGeometry}c presents capacitance data taken from a small dot in a dilution refrigerator with a base temperature of $\SI{45}{\milli\kelvin}$. These data show electron additions from $N=1$ to $N=20$ under an external perpendicular magnetic field, displaying features that fit with a model of a small quantum dot with a parabolic confining potential\cite{fock_bemerkung_1928,darwin_diamagnetism_1931}. As the gate voltage is swept on the $y$-axis, capacitance peaks appear for every electron that enters the dot. Note that the lateral extent of electrons in the dot gradually increases with electron number, decreasing the observed spacing\cite{ashoori_n-electron_1993}. The first dozen electron additions can be grouped into pairs, with the two additions in a given pair showing very similar behavior in magnetic field, in agreement with a simple model in which electron states are occupied by two electrons with opposite spin states. The blue line shows the density and field at which all electrons occupy spin-degenerate states belonging to the lowest orbital Landau level ($\nu=2$ in a 2DES).

The ``zig-zags'' in the evolution of the single-particle peaks with magnetic field indicate crossings of energies of single-particle states. As the magnetic field increases, the energies of the edge states move down relative to those of the bulk states since their orbital magnetic moment is aligned with the field\cite{ashoori_electrons_1996}. At the crossovers of different states, the descending energy of an electronic edge state falls below the ascending energy of a filled bulk state. Consequently, the peak position is expected to zig and zag as the highest-energy electron in the dot moves from one state to another. The zig-zag behavior ends when all electrons fall into the lowest Landau level, at filling factor $\nu=2$. 
\begin{figure}
    \centering
    \includegraphics[width=0.7\textwidth]{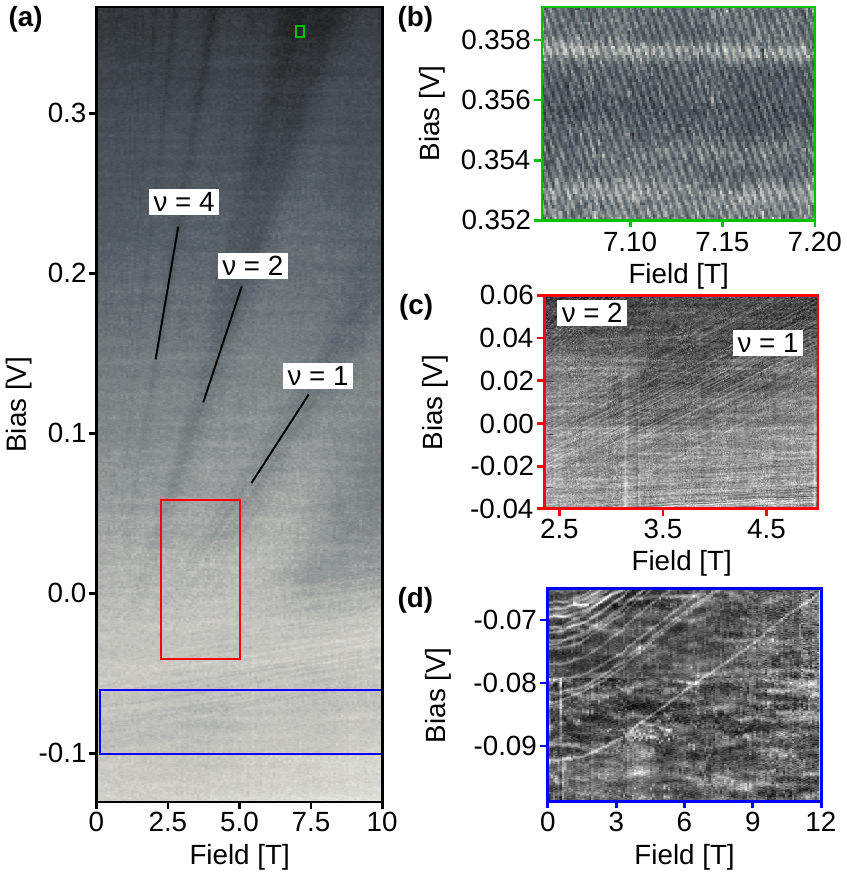}
    \caption{Electron additions in a large 2D quantum dot. (\textbf{a}) Under a perpendicular magnetic field, the 2DES forms Landau levels. We observe dips in the capacitance corresponding to each integer filling factor up to $\nu=8$.  This dot has a lithographic diameter of $\SI{0.8}{\micro\meter}$, and there are roughly two thousand electron additions in the voltage range shown. (\textbf{b}) Between filling factors $\nu=1$ and $\nu=2$, we observe, almost exclusively, states that move down in energy with magnetic field, as expected for edge states. For fixed gate voltage, these appear periodically in magnetic field with a period that corresponds to a flux quantum of $h/e$ threading the dot. However, between filling factors $\nu=2$ and $\nu=5$, the field periodicity of the peaks halves, resulting in $h/2e$ oscillations. (\textbf{c}) Localized states, confined to a small region within the Landau gap, follow the underlying filling factor. The figure shows localized states within the $\nu=1$ gap. The states that are nearly parallel to the field axis are the first electrons in the dot. (\textbf{d}) Capacitance peaks of the first few electrons added to the dot as a function of magnetic field.}
    \label{fig:landauFan}
\end{figure}

Increasing the dot area by an order of magnitude, we observe effectively quantum Hall physics in comparison to ``artificial atom'' physics\cite{ashoori_electrons_1996,kouwenhoven_quantum_1998} that we have shown in Fig.\ \ref{fig:capacitanceGeometry}. For large dot sizes, self-consistent calculations\cite{chklovskii_electrostatics_1992,fogler_chemical_1994} (see supplement) show that the electron density remains nearly uniform over most of the interior of the dot and only diminishes near the dot edges. Under the presence of an external magnetic field, electrons in the mini-2DES develop Landau levels. Fig.\ \ref{fig:landauFan} shows capacitance data from a large dot with a lithographic diameter of $\SI{800}{\nano\meter}$. Fig.\ \ref{fig:landauFan}a shows data from rastering the gate voltage scan and stepping through a wide range of magnetic fields and subtracting offsets from drifts in the measurement. Electrons start to populate the dot at around $V_\text{gate}=\SI{-0.1}{\volt}$. In Fig.\ \ref{fig:landauFan}a, the darker regions correspond to fully filled incompressible states whereas the lighter regions correspond to partially filled compressible states. The data display a clear ``Landau fan,'' indicated by black lines.

In Fig.\ \ref{fig:landauFan}d, the first electron addition to the dot appears at $\SI{-0.094}{\volt}$ at zero magnetic field. The positions of all electron addition peaks evolve with nearly zero slope at small magnetic field. The slope increases with field until the traces appear as nearly straight lines at high fields with slope $\hbar\omega_c/2B$ (where $\omega_c$ is the cyclotron frequency) as would be expected for spatially isolated localized states with a diamagnetic shift due to applied magnetic field\cite{ashoori_single-electron_1992}.  The data are taken with a small $\SI{200}{\micro\volt}$ rms excitation at $\SI{247}{\kilo\hertz}$, a frequency sufficiently lower than the tunneling rate so that the electrons tunnel in and out of the dot in phase with the applied excitation. 

As more electrons are added to the dot, the charges from different localized states merge to create a single small ``droplet'' 2DES. As the droplet expands laterally, its capacitance to the surroundings increases, decreasing the charging energy for subsequent electron additions until we lose the ability to resolve individual capacitance peaks around zero B-field.

Fig.\ \ref{fig:landauFan} shows two distinct groups of interaction-driven localized states that appear at filling factors $\nu=1$ and $\nu=2$. These localized states follow the slopes of the underlying Landau levels and appear only when the Landau level is fully filled. Screening of the electrostatic potential arising from disorder depends sensitively on the Landau level filling factor\cite{chklovskii_electrostatics_1992,efros_non-linear_1988,polyakov_activated_1994}; a partially filled level significantly screens the disorder potential but, around integer $\nu$, the electronic density of states is small and the disorder potential is poorly screened, leading to compressible charge pockets separated by incompressible barriers. In Fig.\ \ref{fig:landauFan}c, the peaks that run parallel to integer $\nu$ arise from electron additions to these pockets\cite{chklovskii_electrostatics_1992,ilani_microscopic_2004,cooper_coulomb_1993}.

Similar peaks appear around every well-developed integer quantum Hall state as also appeared in previous work studying a 2DES gated by a local scanning single electron transistor (SET)\cite{ilani_microscopic_2004}. The charging peaks in that work arose from electrons moving laterally within a very large 2DES to fill individual localized compressible islands in an otherwise incompressible region. In contrast, charge quantization always occurs in our lithographically defined quantum dots, allowing us to observe single electron additions into both compressible and incompressible regimes. This capability led us to observe a remarkable series of periodic electron additions that appear in the compressible regions.

We now focus on performing fine measurements in compressible regions with very large numbers ($\sim2000$) of electrons in the dot. In Fig.\ \ref{fig:landauFan}b, we observe unpredicted states that are evenly spaced in magnetic field. Unlike the electron additions in a small quantum dot (Fig.\ \ref{fig:capacitanceGeometry}c) where the electron addition peaks show zig-zags, the spectra show only uniform straight lines. The energies of these states all move down with increasing magnetic field, as would be expected only for electronic states at the edge of the mini-2DES\cite{ashoori_electrons_1996}. This perfect periodicity and downward movement of all observed states suggest that the edge of the 2DES remains ``compact,'' with all angular momentum states filled. This is the situation that is sometimes expected at filling factor $\nu=1$\cite{oosterkamp_maximum-density_1999,klein_exchange_1995,macdonald_quantum_1993} in a ``maximum density droplet''\cite{macdonald_quantum_1993} (MDD) with no ``edge state reconstructions''\cite{chamon_sharp_1994} occuring in the range of these data sets. In the case of such a $\nu=1$ MDD, one would expect a single electron addition for each additional magnetic flux quantum $h/e$ threading the dot. We note that our data at these densities and filling factors do not display all electron additions to the dot. In a large-area 2D electron system, tunneling from a 3D electrode into the bulk of the 2D system is suppressed exponentially by a magnetic field-induced Coulomb gap\cite{ashoori_equilibrium_1990,ashoori_energy_1993}. In contrast, there is only power-law suppression for tunneling into edge states\cite{chang_chiral_2003}, and electrons tunneling to the edge still do so at short time scales compared to the inverse frequencies of the AC excitations in our measurements. At fixed gate bias, electrons enter the dot as we increase magnetic field, but charge balance can be maintained as electrons not visible to the experiment tunnel out of the center of the dot to the tunneling electrode (see supplement for simulations showing the small variation of the total number of electrons in the dot with varying magnetic field).
\begin{figure}
    \centering
    \includegraphics[width=0.7\textwidth]{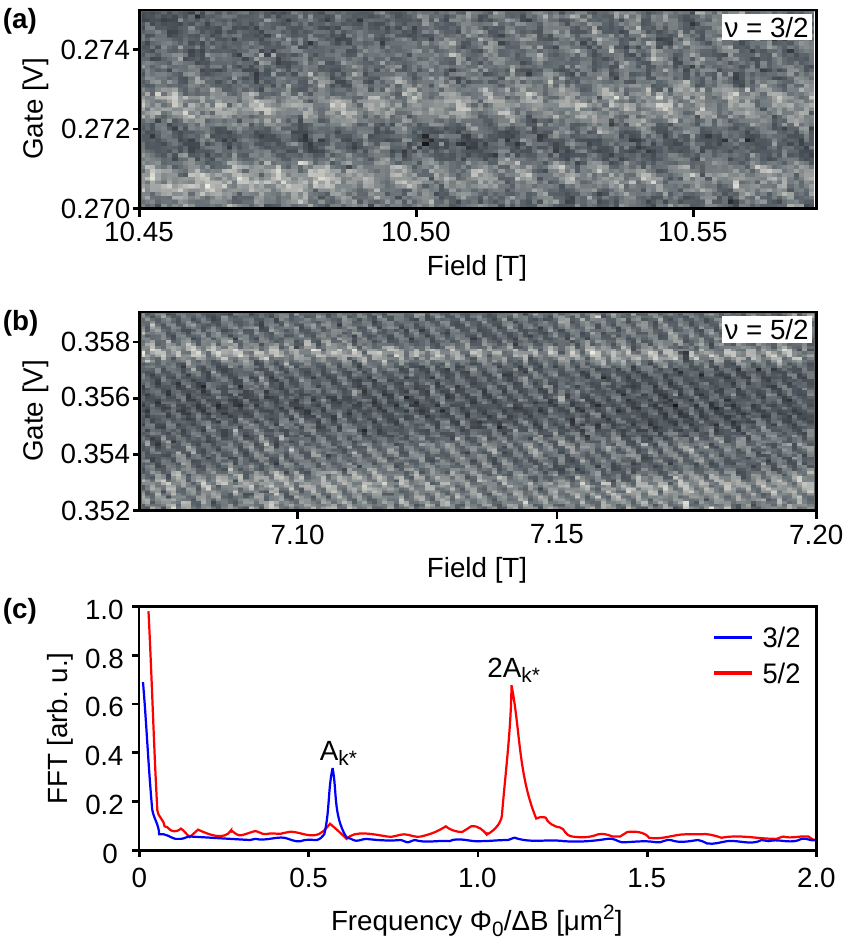}
    \caption{Electron addition spectra in the $N=0$ versus the $N=1$ Landau level. (\textbf{a}) Capacitance data taken as a function of gate voltage and magnetic field in the $\Phi_0=h/e$ regime. The filling factor is near $\nu=3/2$. (\textbf{b}) When the gate voltage is tuned to near filling factor $\nu=5/2$, the periodicity in magnetic field is halved. (\textbf{c}) Corresponding Fourier transforms of two different regions as a function of $\Phi_0/\Delta B$. The blue and red curves are the Fourier transforms near $\nu=3/2$ and $\nu=5/2$, respectively.}
    \label{fig:fourierAnalysis}
\end{figure}

In Fig.\ \ref{fig:fourierAnalysis}, we plot the comparison between the edge states near $\nu=3/2$ and those near $\nu=5/2$. Edge state lines correspond to constant flux, with one flux quantum difference between each constant flux line. The periodicity of electron peaks in field and gate voltage is halved at $\nu=5/2$ as compared to $\nu=3/2$. To illustrate this, we performed Fourier analysis of two regions, plotted in Fig.\ \ref{fig:fourierAnalysis}c. The periodicity in magnetic field when the filling factor is $\nu<2$ (blue curve) is $\Delta B=\SI{13}{\milli\tesla}$. This yields a dot area of $\SI{0.55}{\micro\meter\tothe{2}}$, close to what is expected in our simulations (see supplement). However, when the filling factor is tuned to near $\nu=5/2$ (red curve), we observe a doubling of the electron peak frequency compared to the situation at $\nu=3/2$.

\begin{figure}
    \centering
    \includegraphics[width=0.7\textwidth]{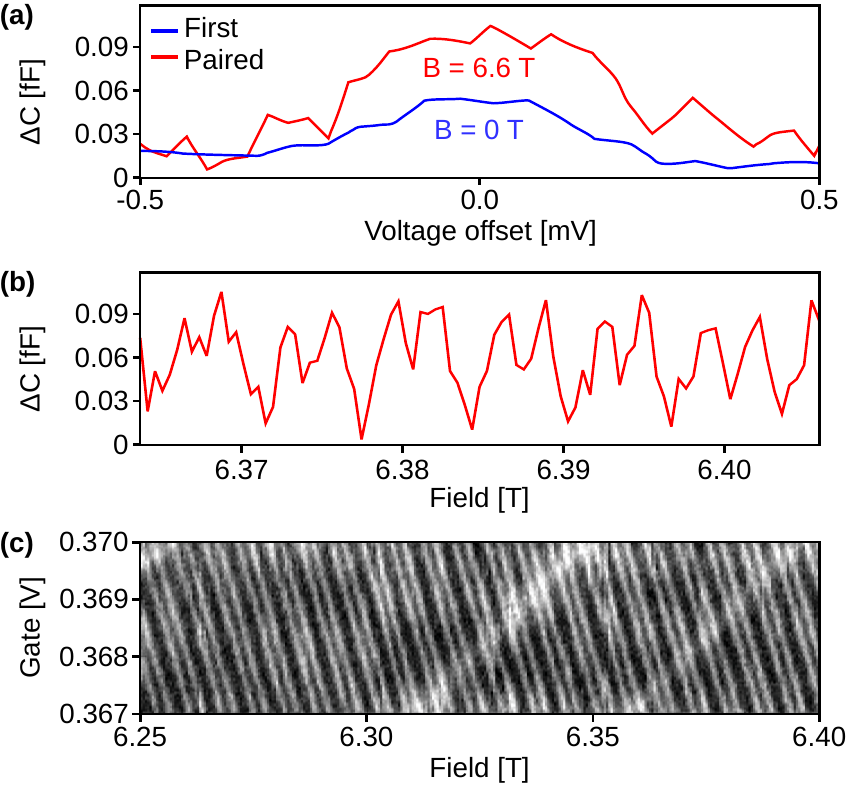}
    \caption{Comparison of individual charging peaks in N=0 and N=1 Landau levels and bunching of paired peaks. (\textbf{a}) Electron charge comparison between the first electrons and the paired electrons. The blue curve (first electrons) is centered at a bias voltage of $\SI{-0.0795}{\volt}$, and the red curve (paired electrons) at $\SI{0.3655}{\volt}$. (\textbf{b}) A line cut along constant bias where the bunching of paired electrons is observed. (\textbf{c}) High magnetic field resolution measurement showing bunched pairs. The filling factor ranges from $\nu=2.9$ at $\SI{6.25}{\tesla}$ to $\nu=2.83$ at $\SI{6.40}{\tesla}$. The AC excitation is $\SI{100}{\micro\volt}$ rms. As this scan is close to $\nu=3$, notice that two localized states appear with positive slope. However, the localized states do not appear to interact with the edge states at all.}
    \label{fig:bunchingOfPairs}
\end{figure}

In capacitance measurements, the capacitance peak height results from the amount of tunneling charge\cite{ashoori_single-electron_1992}. In Fig.\ \ref{fig:bunchingOfPairs}a, we compare the peaks of isolated first electrons (blue curve) to those of edge states near filling factor $\nu=5/2$ (red curve) and find, remarkably, that they arise from the charge of two electrons tunneling back and forth across the barrier (see supplement). Fig.\ \ref{fig:bunchingOfPairs}b is a line cut from Fig.\ \ref{fig:bunchingOfPairs}c demonstrating that $2e$ charge transfer happens in the $h/2e$ regime. We observe this behavior when the density is tuned between filling factor $\nu=2$ and $\nu=5$. When the lowest ($N=0$) Landau level is partially filled, Coulomb repulsion dominates.

These double-electron additions contradict the prediction of Coulomb blockade theory\cite{beenakker_theory_1991,silsbee_comment_1990,devoret_introduction_1992} that more energy is required to add each successive electron to a quantum dot as a result of electron repulsion. While prior measurements on more disordered dots at lower densities also showed a violation of Coulomb blockade with pairing and bunching of electron addition peaks\cite{ashoori_single-electron_1992,zhitenev_localization-delocalization_1999,zhitenev_periodic_1997,brodsky_localization_2000}, the current results differ in that \textit{all} observed electron additions, over nearly the entire range of $\nu$ between integer values of $\nu$, occur as pairs. They thus appear more similar to pairing phenomena observed in charging superconducting islands\cite{lafarge_two-electron_1993,tuominen_experimental_1992,joyez_observation_1994}. Due to the superconducting gap and pairing of electrons in the condensate, ground states of even and odd numbers of electrons have energies differing by the superconducting gap, resulting in parity-induced suppression of Coulomb blockade\cite{matveev_parity-induced_1993,hekking_coulomb_1993}.

Another surprising observation is that the double-electron peaks themselves bunch together at all filling factors between $\nu=2$ and $\nu=5$ with the only clear exception at $\nu=5/2$, where the peaks are evenly spaced. Beyond $\nu=5$, the peaks become difficult to discern. Figs.\ \ref{fig:bunchingOfPairs}b and \ref{fig:bunchingOfPairs}c demonstrate this bunching, presenting data taken with a $\SI{100}{\micro\volt}$ rms excitation. In the supplement, we provide data similar to that shown in Fig.\ \ref{fig:bunchingOfPairs}c but over a broad filling factor range including $\nu=5/2$. Similar behaviors, such as the $h/2e$ periodicity, bunching phenomena, and localized states, all occur in another dot with a lithographic area of $\SI{0.325}{\micro\meter\tothe{2}}$.

The range of filling factors over which pairing and bunching take place suggests a relation to pairing phenomena observed in an entirely different type of experiment\cite{choi_robust_2015,sivan_interaction-induced_2018} performed on an electronic version of a Fabry-Perot interferometer (FPI). The FPIs are fabricated on a 2DES confined by two quantum point contacts. Allowing only the outermost edge channel through the QPCs, the presence of electrons as the fundamental unit of charge suggest that conductance oscillations through the FPI should occur with period $h/e$. The FPI experiment displays this predicted behavior between filling factors $\nu=1$ and $\nu=2$. However, the periodicity in magnetic field is halved between filling factors $\nu=5/2$ and $\nu=5$, yielding $h/2e$ oscillations (in contrast with $2e$ flux quantum between $\nu=2$ and $\nu=5$ in our experiment), with quantum shot noise measurements also suggesting a quasiparticle charge of $2e$ rather than $e$ (as seen at lower filling factors) in this regime\cite{choi_robust_2015}.  We note that shot noise measurements result from out-of-equilibrium tunneling events, while the two-electron additions in our experiment, existing in the limit of small excitation drive, take place in thermodynamic equilibrium. Moreover, while the FPI data reveal a mechanism that binds two electrons together as they encircle the FPI, our results now show that the energetics of this binding is sufficiently strong to produce violation of Coulomb blockade and also suggest the presence of coherent tunneling of the two electrons through a thick tunnel barrier, as discussed below. Finally, our data (see supplement and Figs. S8 and S9) appear to demonstrate that each of the two electrons in a pairing enters a single edge state, and alternate pairs in a two-pair bunch enter different edge states.

The observed tunneling into the edge states shows no sign of decreased tunneling rate that would appear as phase shifts in the charging response. ``Negative-$U$'' models involving sequential tunneling and rearrangements of the electronic system\cite{raikh_two-electron_1996,hamo_electron_2016} that place two peaks at the same gate voltage would lead to strongly suppressed tunneling rates that we do not observe. To have two electrons tunneling back and forth between the dot and the tunneling electrode at the exact same gate voltage, the first electron in a pair must first tunnel into the dot with less than the required energy $\Delta E$ to produce the rearrangement (where the rearrangement also produces shifts in single electron peak positions comparable to $\Delta E$). Therefore the second electron in the pair must tunnel into the dot at a rate that is fast compared to $\Delta E/\hbar$ ($\sim\SI{e11}{\per\second}$ for the $\approx\SI{0.4}{\milli\electronvolt}$ energy barrier as seen from the gate voltage spacing of single electron peaks). As tunneling rates in our dots from the tunneling electrode are on the order of $\SI{e6}{\per\second}$, such negative-$U$ models cannot explain our data. Another possibility would then be an effective zero repulsion between electrons in a pair, but we know of no model for this. The answer may lie in coherent tunneling of the two electrons. For instance, in tunneling to superconducting islands from a normal metal electrode\cite{joyez_observation_1994}, the tunneling rate is enhanced due to Andreev processes compared with the suppressed incoherent tunneling of two electrons\cite{hekking_coulomb_1993}. Indeed, the occupancy of uniform (unbunched) pairs occurring only very near $\nu=5/2$ may suggest such coherent tunneling could have a connection with the 5/2 fractional quantum Hall state where theory describes a potential Cooper pairing of composite fermions\cite{scarola_cooper_2000}.

\begin{acknowledgments}
Measurement and analysis of quantum dot data were carried out with support from BES Program of the Office of Science of the US DOE, contract no. FG02-08ER46514. Fabrication of the quantum dots was performed at MIT Microsystems Technology Laboratory with support from the Gordon and Betty Moore Foundation, through grant GBMF2931 and the STC Center for Integrated Quantum Materials, NSF Grant No. DMR-1231319. S. Aronson was supported by the National Science Foundation Graduate Research Fellowship under grant no. 1122374. We thank L. Levitov, P.A. Lee, and L. Glazman for helpful discussions.
\end{acknowledgments}

\bibliography{main}

\renewcommand{\thefigure}{S\arabic{figure}}
\setcounter{figure}{0} 

\section{Quantum dot geometry and growth}

\begin{table}[htb]
    \centering
    \begin{tabular*}{6in}{@{\extracolsep{\fill}}lllr}
        & substrate & GaAs (undoped) &\\
        \hline
        1  & bottom initial growth & GaAs (intrinsic) & $\SI{3000}{\angstrom}$\\
        2  & bottom blocking barrier & Al${}_{0.323}$Ga${}_{0.677}$As & $\SI{4000}{\angstrom}$\\
        3  & bottom electrode & GaAs (n+, $4\times 10^{18}$) & $\SI{2000}{\angstrom}$\\
        4  & bottom electrode & GaAs (n+, $1\times 10^{18}$) & $\SI{1000}{\angstrom}$\\ 
        5  & bottom spacer & GaAs (intrinsic) & $\SI{30}{\angstrom}$\\ 
        6  & cold growth/diffusion barrier & GaAs (intrinsic) & $\SI{30}{\angstrom}$\\
        \hline 
        7  & blocking barrier & Al${}_{0.323}$Ga${}_{0.677}$As (intrinsic) & $\SI{600}{\angstrom}$\\ 
        \hline 
        8  & quantum well  & GaAs (intrinsic) & $\SI{230}{\angstrom}$\\
        \hline 
        9  & tunnel barrier & Al${}_{0.323}$Ga${}_{0.677}$As (intrinsic) & $\SI{90}{\angstrom}$\\ 
        \hline 
        10   &  top spacer  & GaAs (intrinsic) & $\SI{25}{\angstrom}$\\
        11   & top electrode                & GaAs (n+, $1\times 10^{18}$) & $\SI{200}{\angstrom}$\\ 
        12   & top electrode                & GaAs (n+, $4\times 10^{18}$) & $\SI{400}{\angstrom}$\\ 
        13   & top electrode                & Delta doped top layers & $\SI{25}{\angstrom}\times8$\\ 
        \hline 
        & top surface & &
    \end{tabular*}
    \caption{The growth sheet for the GaAs/AlGaAs wafer used to fabricate the quantum dots.}
    \label{tab:mbe}
\end{table}

\begin{figure}[htb]
    \centering
    \includegraphics[width=0.85\textwidth]{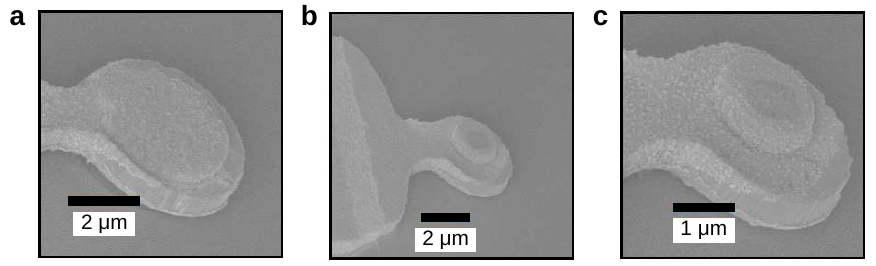}
    \caption{SEM images of completed devices.}
    \label{fig:semTopContact}
\end{figure}

\section{Measurement scheme}
In order to sense tiny capacitance changes, we have use an AC charge sensing technique based on a capacitance "bridge on a chip"\cite{ashoori_single-electron_1992}. In this technique, we utilize the bridge to balance the sample impedance against a known reference capacitor by applying excitation and reference voltages that are 180 degrees out of phase. The voltage at the balance point is nulled when
\begin{equation}
\label{eqn:balance}
    \frac{\tilde{V}_{\text{exc}}}{Z_{\text{sample}}}+\frac{\tilde{V}_{\text{ref}}}{Z_{\text{ref}}}=0
\end{equation}
where $Z_{\text{ref}}=1/i\omega C_{\text{ref}}$ is the impedance of the reference capacitor and $Z_{\text{sample}}$ is the sample impedance. To make the reference capacitor, we deposit metal lines on a silicon chip, yielding a capacitance of about $\SI{40}{\femto\farad}$. If the sample is purely capacitive, the balance condition becomes
\begin{equation}
\label{eqn:capacitiveBalance}
    C_{\text{sample}}=C_{\text{ref}}\frac{\tilde{V}_{\text{ref}}}{\tilde{V}_{\text{exc}}}
\end{equation}

\begin{figure}
    \centering
    \includegraphics[width=0.85\textwidth]{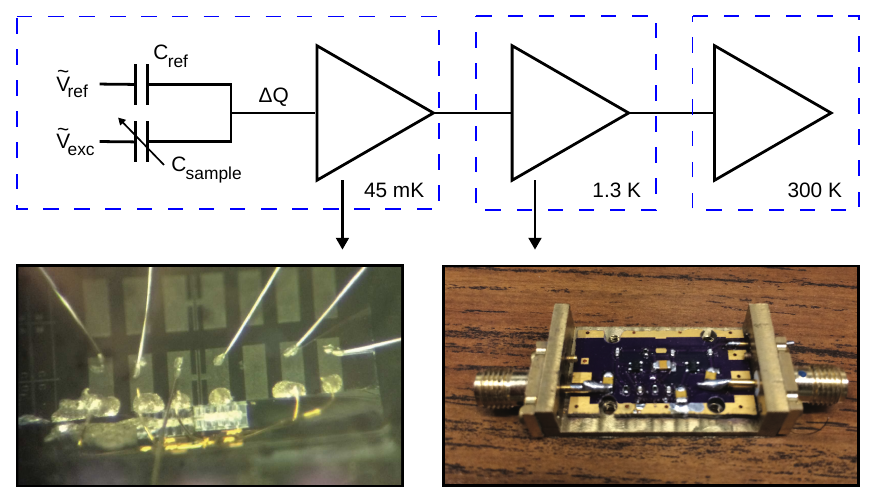}
    \caption{Measurement scheme from cryogenic temperatures to room temperature.
The image on the left shows how cleaved transistors are mounted on the edge of a silicon chip. The image on the right shows the $\SI{1}{\kelvin}$ pot amplifier.  
}
    \label{fig:measurementScheme}
\end{figure}

The main motivation behind using this technique is to eliminate the effect of constant stray capacitance. Wire bonds and metallic pads deposited during the fabrication process produce a stray capacitance from the bridge balance point to ground of about $\SI{150}{\femto\farad}$. The capacitance bridge allows for a precise measurement of the sample capacitance without needing an independent measurement of the shunt capacitance or knowledge of the gain of any amplifiers.

To balance the bridge, we apply two different excitations to the reference capacitor while applying a known, fixed excitation to the sample. Measuring the voltage at the balance point for each case, we then compute the excitation voltage that satisfies the balance condition. The changes in capacitance are often very small compared to the total sample capacitance. When this is the case, we do not need to re-balance the bridge at each point. Instead, we can measure the out-of-balance signal directly and convert that to capacitance.

\section{Electron temperature}
During one cycle of the excitation, an electron can tunnel back and forth across the barrier many times. We choose an AC frequency much lower than the roll-off frequency of the tunnel barrier so that we operate in the low-frequency limit. For instance, in the zero temperature limit, the quantum level in the dot tends to be occupied during positive portions of the AC excitation and unoccupied during negative portions, creating a square wave signal at the balance point. Depending on how far the DC voltage is from the center of the peak, what we measure at the lock-in is the variable duty cycle square wave amplified through the cold amplifiers. The lock-in amplifier picks up the Fourier coefficient of this voltage at the excitation frequency $\omega$. 

Consider a DC gate voltage $V_0$ to be set at the center of an isolated capacitance peak. Then, add a small excitation $V_{\text{exc}}= v \cos{\omega t}$ so that the net voltage across the tunnel capacitor becomes $V_{\text{sample}}(t)=V_0 + v \cos{\omega t}$. The occupation probability of that single electron level is given by the Fermi distribution $f(V_{\text{sample}})$. The voltage change at the balance point due to a tunneling event is thus the Fermi distribution function multiplied with $e/\eta C_{\text{shunt}}$, where $\eta$ is the geometric ``lever-arm'' that is the ratio of the distance from the tunneling electrode to the gate electrode to the distance from the tunneling electrode to the quantum dot. For the dots used in this experiment, $\eta=3.7$, as determined by comparing the measured capacitance of depleted quantum wells to that of filled wells. The signal measured at the lock-in amplifier is 
\begin{equation}
V_{\text{lockin}}(V_0) = A_v\frac{\omega e}{\sqrt{2}\pi \eta C_{\text{shunt}}} \int_0^{2 \pi / \omega} f(V_0+v \cos{\omega t}) \cos{\omega t}\,\,dt
\end{equation}
where $A_v$ is the total gain from the balance point to the input of the lock-in. The division by $\sqrt{2}$ is made because the lock-in reads the rms amplitude of the Fourier coefficient. A change of variables, $U=V_0+v \cos(\omega t)$, and integration by parts yields
\begin{equation}
V_{\text{lockin}}(V_0) = A_v \frac{\sqrt{2}e}{\pi \eta C_{\text{shunt}}} \int_{V_0 - v}^{V_0 + v} \frac{df(U)}{dU} \sqrt{1-\left(\frac{V_0-U}{v}\right)^2}\,\,dU
\end{equation}

At zero temperature, the derivative $df(U)/dU$ is simply given by $\delta\left(U-V_n\right)$, where $V_n$ is the gate voltage at which the $n$th electron enters the dot. The expected response at the lock-in as a function of gate voltage is half an ellipse of height $\sqrt{2}A_ve/\pi\eta C_\text{shunt}$ with base width of twice the amplitude of the AC excitation, centered at the voltage $V_n$\cite{ashoori_single-electron_1992}. At nonzero temperatures, and for $v <\eta k_B T/e$, the height of the peak drops and becomes amplitude dependent. For $v \gg \eta k_B T/e$, the peak height saturates at $\sqrt{2}A_ve/\pi\eta C_\text{shunt}$, and only its
width increases.
\begin{figure}
    \centering
    \includegraphics[width=0.85\textwidth]{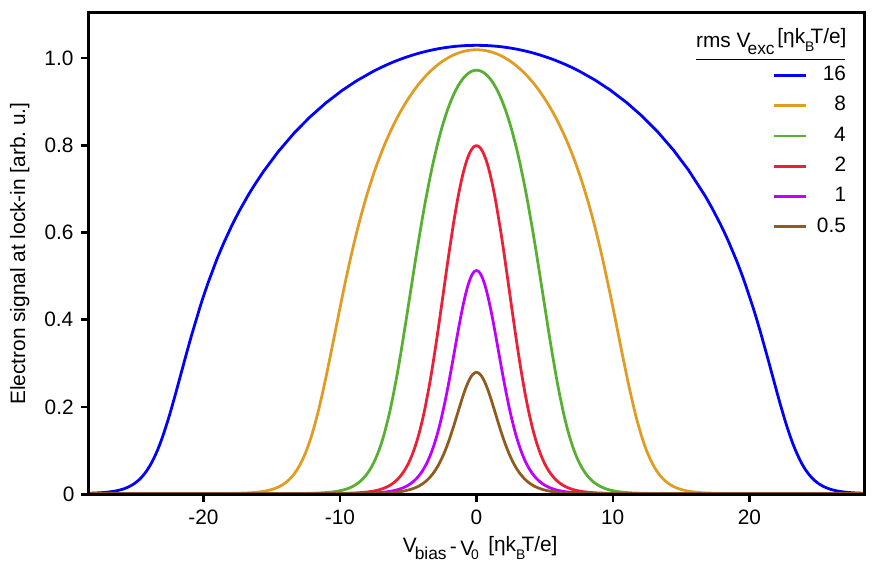}
    \caption{Simulation of the
voltage response of an electron addition to AC excitations of varying amplitude}
    \label{fig:electronTempSimulation}
\end{figure}

In order to understand this behavior, in Fig.\ \ref{fig:electronTempSimulation}, we have simulated the voltage response of an electron addition to AC excitations of
varying amplitude. On the $y$-axis we present the normalized response at the lock-in with respect to the maximum response $\sqrt{2}A_ve/\pi\eta C_\text{shunt}$. The rms amplitude of the AC excitation is varied from $0.5 \eta k_B T/e$ to $16 \eta k_B T/e$, increasing by a factor of two in each subsequent curve. On the $x$-axis, we sweep the applied DC voltage around the center of the peak $V_0$.

Fig.\ \ref{fig:electronTempSimulation} shows that for $v_{\text{rms}} \gg \eta k_B T/e$, the peak height saturates and only its width increases. For $v_{\text{rms}} <4\eta k_B T/e$, the height of the peak drops and becomes amplitude dependent. The full width at half maximum height stays constant at roughly $4\eta k_B T/e$. As is clearly shown in Fig.\ \ref{fig:electronTempSimulation}, to get the maximum voltage response, we need to drive the tunnel capacitor with an AC excitation greater than $v_{\text{rms}} = 4\eta k_B T/e$. Above this excitation voltage, the peaks broaden. However, below this excitation voltage, we will lose the signal height.
\begin{figure}
    \centering
    \includegraphics[width=0.85\textwidth]{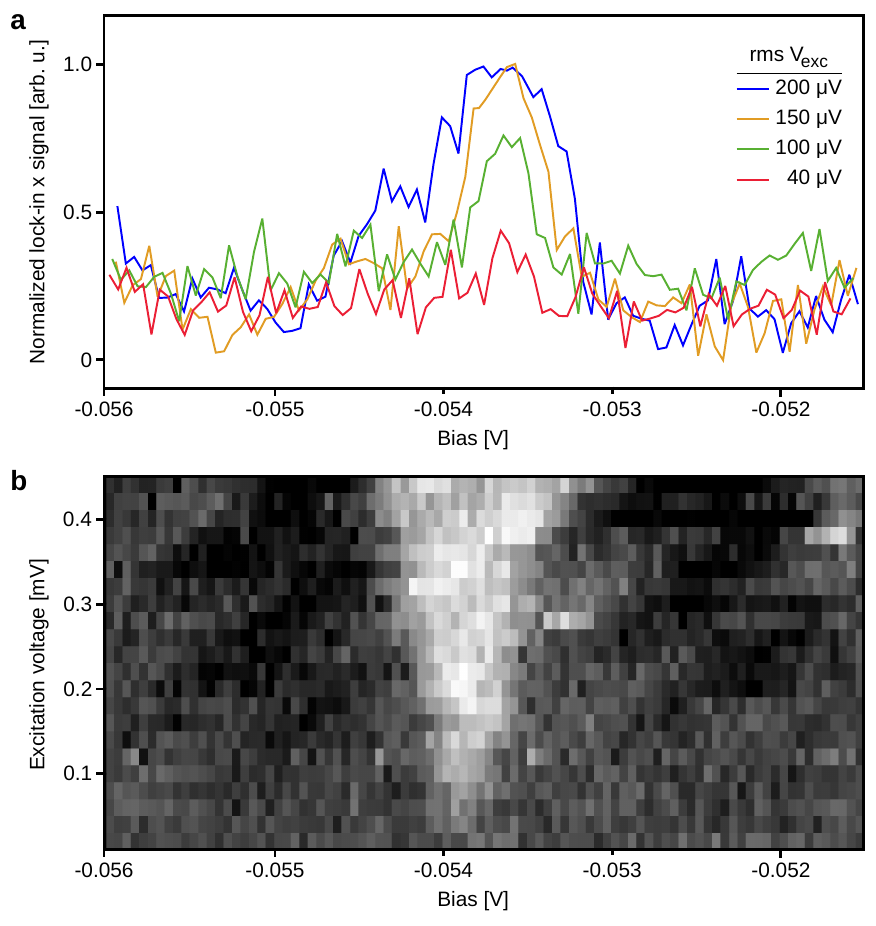}
    \caption{(\textbf{a}) Line cuts from (\textbf{b}), which shows the measured lock-in signal at varying excitation voltage.}
    \label{fig:electronTemp}
\end{figure}

Our experimental data from a single isolated capacitance peak is shown in Fig.\ \ref{fig:electronTemp}. To estimate the electron temperature, we examine two features of the data. The excitation voltage at which the voltage response is significantly lower than the maximum response is roughly $4\eta k_B T/e$. The peak width at the lowest excitation which we can still observe a height is also roughly $4\eta k_B T/e$. Looking at the amplitude that is measured at the lock-in, we observe that below an excitation voltage of $V_{\text{exc}}=\SI{100}{\micro\volt}$, the peak height starts to drop. This gives a temperature estimation of ($4\eta k_B T/e =\SI{100}{\micro\volt}$) $\SI{78}{\milli\kelvin}$. This is slightly warmer than the $\SI{45}{\milli\kelvin}$ base temperature of the dilution refrigerator. The power dissipated by the transistors is likely heating the sample. In Fig.\ \ref{fig:electronTemp}b, we observe that the peak width is linearly increasing with the excitation voltage above $\SI{60}{\micro\volt}$.

\section{Between filling factors $\nu = 1$ and $\nu = 2$}

\begin{figure}
    \centering
    \includegraphics[width=0.85\textwidth]{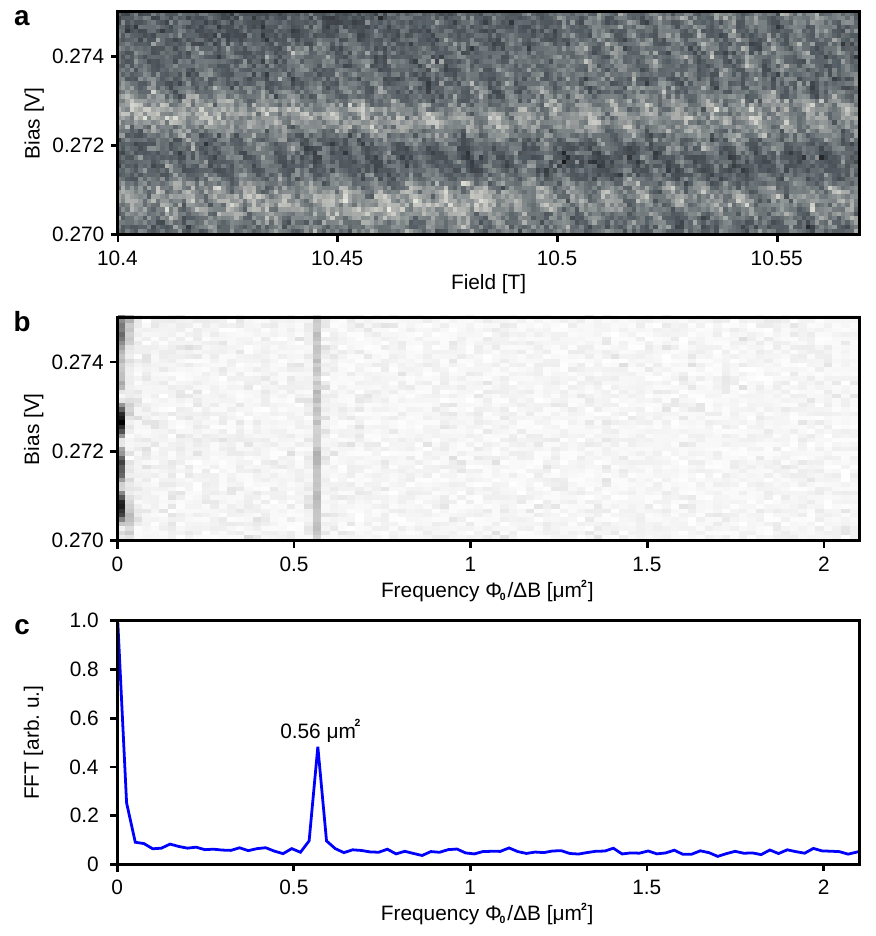}
    \caption{(\textbf{a}) Capacitance peaks in the compressible region between filling factor $\nu=1$ and $\nu=2$. (\textbf{b}) The discrete Fourier transform of field sweeps at each density. (\textbf{c}) The mean Fourier transform indicating the frequency of the single electron peaks.}
    \label{fig:fourierV1}
\end{figure}

To investigate the field dependence of capacitance peaks at constant density, we perform discrete Fourier transforms. Between filling factors $\nu = 1$ and $\nu = 2$, Fig.\ \ref{fig:fourierV1} shows capacitance peaks appearing periodically in magnetic field with a period that corresponds to a flux quantum of $h/e$ threading the dot. The Fourier transform in Fig.\ \ref{fig:fourierV1}c shows that the periodicity in magnetic field when the filling factor is between $\nu = 1$ and $\nu = 2$ is $\Delta B = \SI{7.4}{\milli\tesla}$. We thus deduce an area of $h/e \Delta B = \SI{0.56}{\micro\meter\tothe{2}}$. This area is compatible with simulations for this structure that we describe below.

Our observations were found to be robust and reproducible in different quantum dots of different sizes. In a smaller dot, between $\nu=1$ and $\nu=2$, we observed that capacitance peaks have field periodicity of $\Delta B = \SI{12.8}{\milli\tesla}$, indicating an area of $\SI{0.325 }{\micro\meter\tothe{2}}$. However, between filling factors $\nu = 2$ and $\nu = 5$, the field periodicity of the peaks halves, resulting h/2e oscillations. This is clearly evident in the edge states plot near filling factor  $\nu = 5/2$. The red curve in Fig.\ 3c shows that the Fourier transform has a frequency peak twice as large, indicating multiple windings.

Using the Fourier transform, we can also estimate the capacitance peak height. In this analysis we use the discrete Fourier transform of a sequence of N numbers as
\begin{equation}
X_{k}  =\frac{1}{N}\sum _{j=1}^{N}x_{j}\cdot e^{-{\frac {2\pi i}{N}}kj}
\end{equation}
where $x_j$ is the $j$th component of the sequence. Assume the sequence $x_j$ has the sinusoidal form $x_j = A e^{{\frac {2\pi i}{N}}\bar{k}j}$, with frequency $\bar{k}$. Then, the Fourier sequence will have a peak at $\bar{k}$ with an amplitude proportional to $A$:
\begin{equation}
\begin{split}
X_{k}  & = \frac{1}{N}\sum _{j=1}^{N}A e^{{\frac {2\pi i}{N}}\bar{k}j}  \cdot e^{-{\frac {2\pi i}{N}}kj} \\
& = \frac{1}{N}\sum _{j=1}^{N}A e^{{\frac {2\pi i}{N}}\left(\bar{k}-k\right)j}
\end{split}
\end{equation}
In the limit $N \to \infty$ this sum is non-zero only when $k=\bar{k}$, yielding $X_{\bar{k}}=A$. In other words, the height of the Fourier component is linearly proportional to the capacitance peak height. We can utilize this to investigate electron peak heights when the data are noisy.

\section{Bunching phenomena}

\begin{figure}
    \centering
    \includegraphics[width=0.8\textwidth]{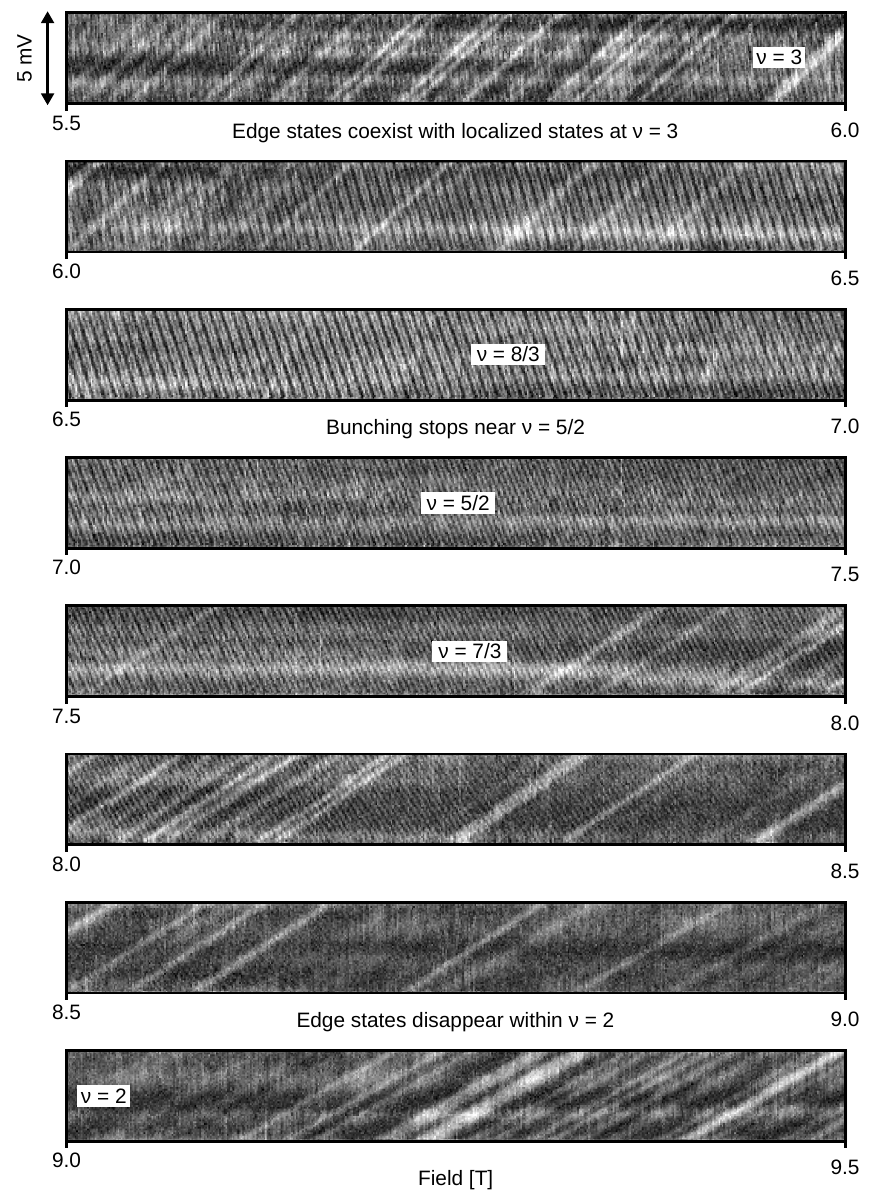}
    \caption{Electron additions over the full range of filling factors from $\nu=3$ to $\nu=2$. Capacitance measurements show a periodicity that corresponds to $h/2e$ oscillations between filling factor $\nu=2$ and $\nu=5$, and the electron additions bunch in pairs in this range except for in the vicinity of $\nu=5/2$ where bunching disappears for a very narrow range. The traces are taken line by line across a narrow gate voltage ($\SI{0.370}{\volt}$ to $\SI{0.375}{\volt}$) at different magnetic fields. The approximate number of electrons in the dot is 2300.}
    \label{fig:wideFieldSweep}
\end{figure}

Fig.\ \ref{fig:wideFieldSweep} shows the bunching of the double-electron peaks seen while sweeping the field from $\nu=3$ to $\nu=2$. The bunching occurs for all compressible filling factors between $\nu=3$ to $\nu=2$ except for a narrow region around $\nu=5/2$. The bunching phenomena persists up to filling factor $\nu=5$, beyond which we lose resolution of individual peaks. The halving of periodicity in magnetic field and associated double-peak height capacitance peaks, the bunching phenomena, and the localized states have also been observed in another sample which has an area of $\SI{0.325}{\micro\meter\tothe{2}}$.

\begin{figure}
    \centering
    \includegraphics[width=0.85\textwidth]{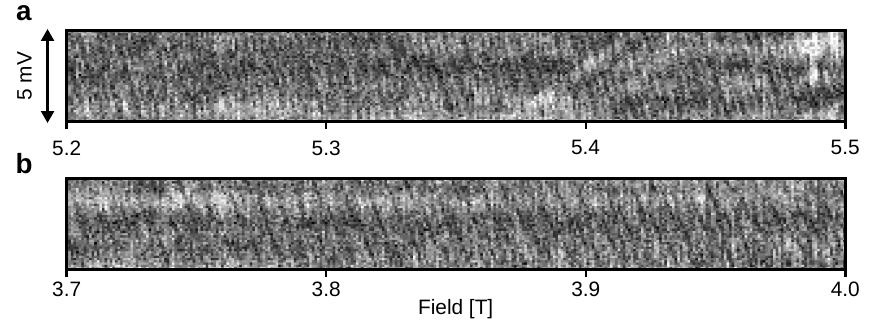}
    \caption{Edge states between (\textbf{a}) $\nu = 3.4$ and $\nu = 3.2$ and between (\textbf{b}) $\nu = 4.8$ and $\nu = 4.4$. Due to insufficient visibility of peaks, we are not able to determine if the double-electron peaks are uniformly spaced or bunched at $\nu=7/2$.}
    \label{fig:beyondnu3}
\end{figure}

The pairing appears between filling factors $\nu = 3$ and $\nu = 5$ except for right where the density is tuned into fully filled incompressible gaps. Within the gaps, we lose the resolution of the edge states mainly because of dominance of the localized states. In Fig.\ \ref{fig:beyondnu3}, we show the capacitance peaks from two different compressible regions. Fourier analysis shows that beyond filling factor $\nu=3$, we have the same magnetic field periodicity as the pairing between $\nu=2$ and $\nu=3$. This rules out the simple argument in which the number of available edge states determines the magnetic field periodicity. One might naively expect that, for instance, between $\nu=3$ and $\nu=4$, we have three edge states producing $h/3e$ oscillations.

\begin{figure}
    \centering
    \includegraphics[width=0.85\textwidth]{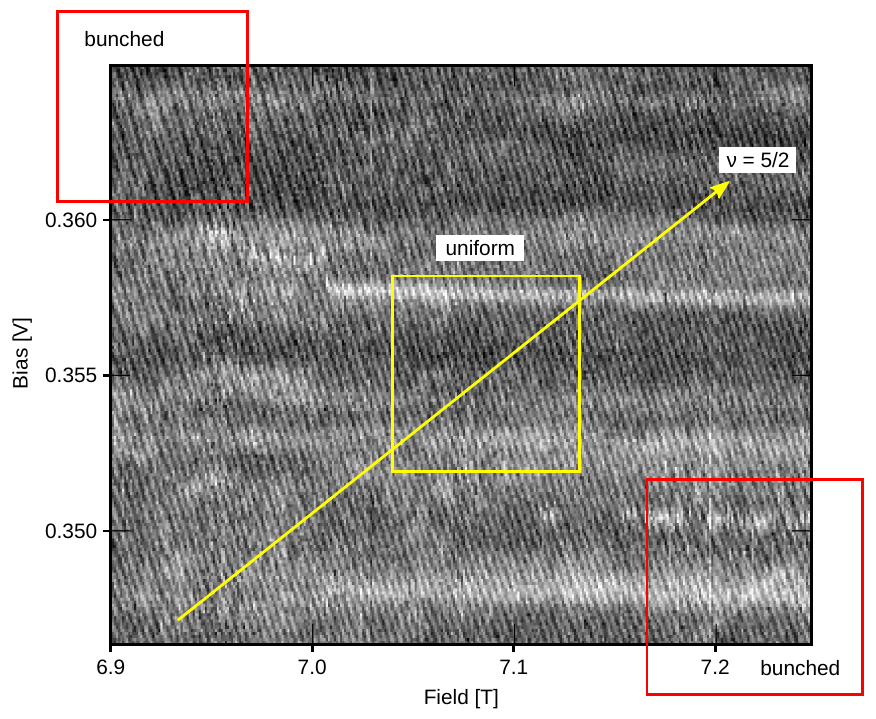}
    \caption{We observe that edge states bunch everywhere between $\nu = 2$ and $\nu = 5$, except for the narrow band at $\nu = 5/2$. This figure illustrates how the bunching transitions to uniform peaks.}
    \label{fig:bunchingTransition}
\end{figure}

\begin{figure}
    \centering
    \includegraphics[width=0.85\textwidth]{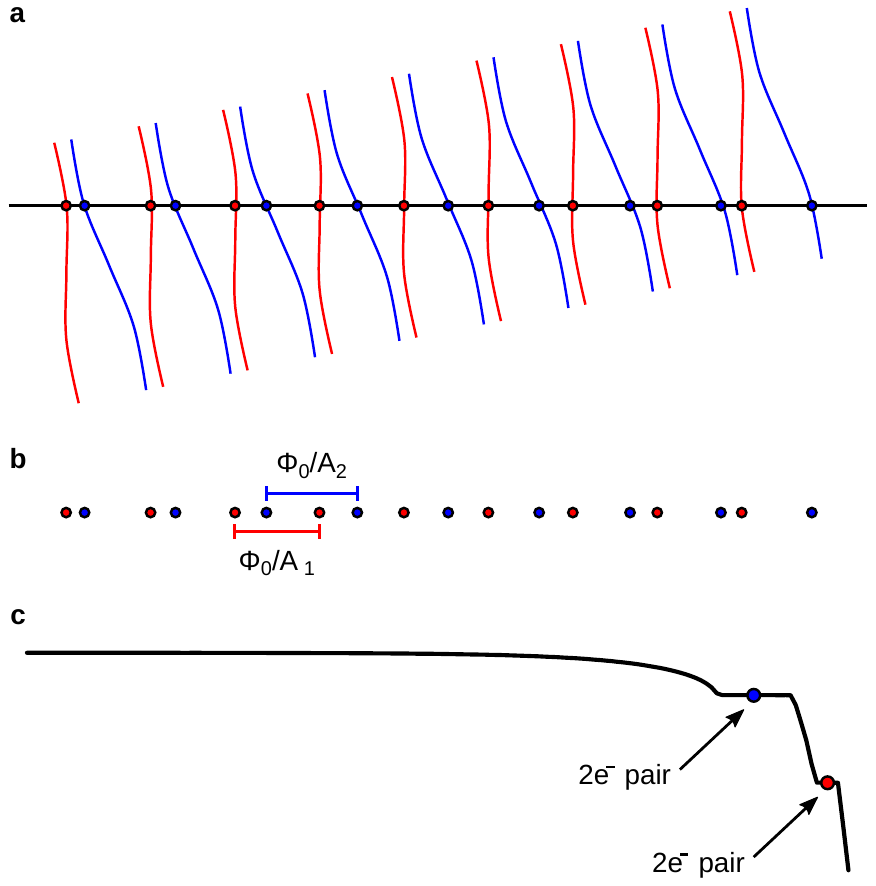}
    \caption{(\textbf{a}) Cartoon showing the evolution of the bunched pairs through $\nu=5/2$. The first paired peak in a bunch in the $\nu>5/2$ region becomes the second paired peak in a bunch in the $\nu<5/2$ region. (\textbf{b}) Peak positions along a cut of constant density. $A_1$ and $A_2$ are the areas enclosed by the two edge states between $\nu=2$ and $\nu=3$. (\textbf{c}) Possible microscopic origin of the beating effect. Both electrons in a paired peak enter the same edge state. Since the two edges enclose different areas, the two sets of paired peaks have slightly different periods.}
    \label{fig:bunchingAnalysis}
\end{figure}

Fig.\ \ref{fig:bunchingTransition} shows high-resolution capacitance data taken around $\nu=5/2$. Paired electron peaks bunch together on one side of $\nu=5/2$, transition to uniform spacing very close to $\nu=5/2$, and transition back to bunching on the other side. Upon careful analysis of the data, we find that any peak that starts as the first peak in a bunch above $\nu=5/2$ becomes the second peak in a bunch below $\nu=5/2$. We illustrate this phenomemon schematically in Fig.\ \ref{fig:bunchingAnalysis}a. Taking a line cut at constant density and varying magnetic field, we obtain two sets of peaks indicated by red and blue circles. We present a simple microscopic picture that may explain this effect: The quantum dot has two quantum Hall edge states near $\nu=5/2$. Both electrons in a given pair enter the same edge state, with the pairs themselves alternating between the outer and inner edge. The two edge states encircle slightly different areas $A_1$ and $A_2$, which correspond to two different field periodicities $\Phi_0/A_1$ and $\Phi_0/A_2$. This leads to beating and bunching of the paired capacitance peaks as shown in Fig.\ \ref{fig:bunchingAnalysis}b. While this picture is consistent with the observed bunching of electron pairs, it does not provide a reason as to why electrons enter the edge states in pairs in the first place. It also does not explain why the peaks become uniformly spaced only very close to $\nu=5/2$.

\FloatBarrier

\section{Simulations}
We have also performed numerical simulations of the quantum dot’s electron density and electrostatic potential landscape to further understand the behavior of the dot as a function of applied bias and field. We model the device on a two-dimensional grid with a spacing of $\SI{15}{\angstrom}$ in both the radial and vertical directions, assuming cylindrical symmetry. The simulation iteratively applies Poisson’s equation to update the potential at each point on the grid, and then updates the charge density in both the dot itself and the GaAs pillar opposite the tunnel barrier. This simulates charge transfer across the barrier. Convergence of the simulation brings the chemical potential of the dot and the pillar into equilibrium. We take into account electrostatic repulsion and quantum level spacing when determining the chemical potential in the dot but ignore exchange interactions between individual electrons. We assume the density of states of the pillar to be infinite. We model band offsets at the interfaces between different materials as constant offsets in the electrostatic potential.
\begin{figure}
    \centering
    \includegraphics[width=\textwidth]{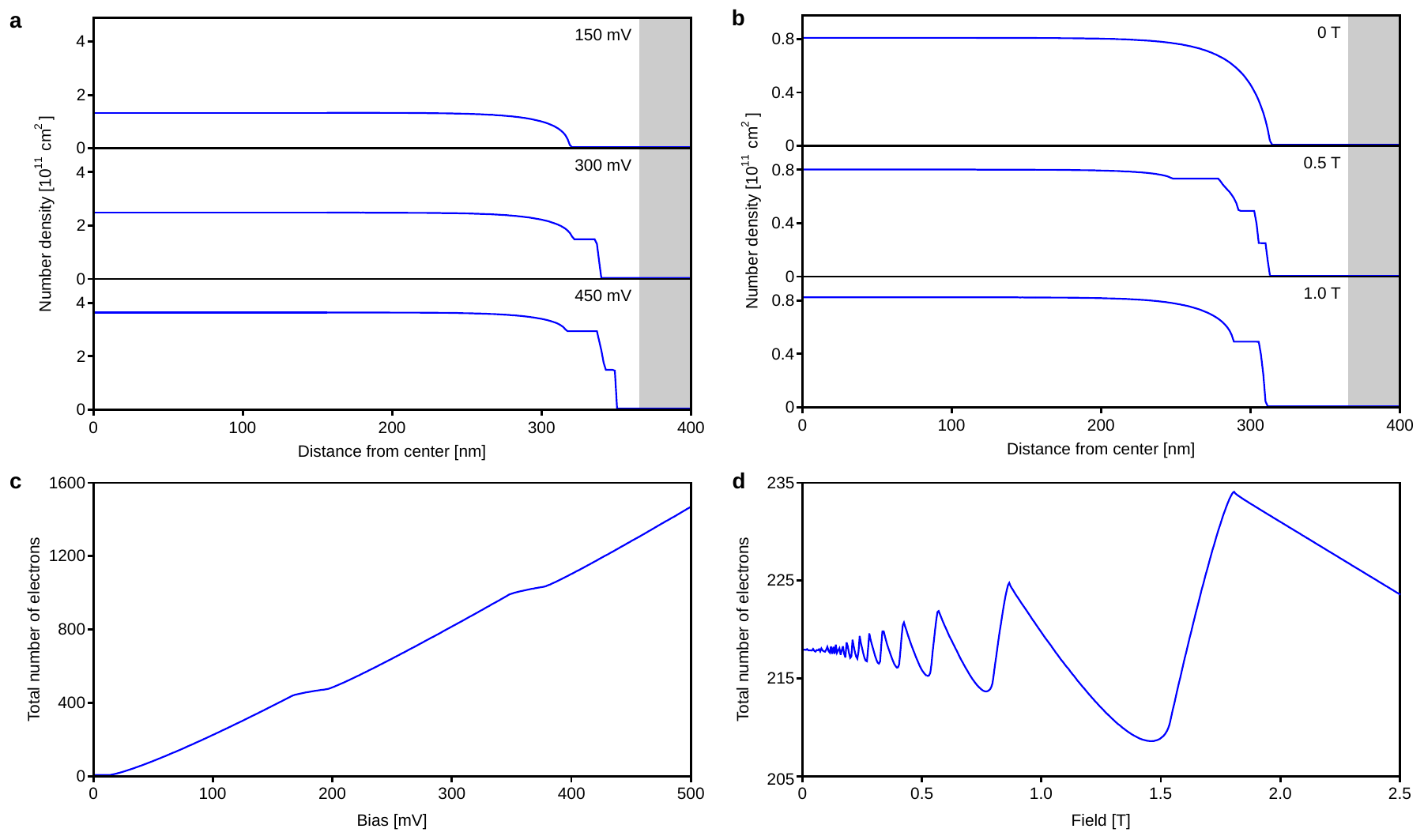}
    \caption{(\textbf{a}) Simulated electron density profiles in the dot at $\SI{3.0}{\tesla}$. The shaded region corresponds to the depletion region in the pillar. (\textbf{b}) Simulated electron density profiles in the dot at a bias voltage of $\SI{100}{\milli\volt}$. (\textbf{c}) Simulated evolution of the total occupancy of the dot with applied bias at $\SI{3.0}{\tesla}$. (\textbf{d}) Simulated evolution of the total occupancy of the dot with magnetic field at $\SI{100}{\milli\volt}$.}
    \label{fig:simulations}
\end{figure}

Our simulations provide information about both the potential landscape of the device and the evolution of the electron density in the dot with applied bias and field. The depletion region on the side of the pillar is about 25 nm across. The diameter of the dot grows to roughly 700 nm as the bias voltage is increased to 400 mV, approaching the 800 nm lithographic diameter of the mesa itself. A zero-field simulation shows that the density of the dot is fairly constant from the center outwards before falling off suddenly near its edge, suggesting that the dot is well-described as a 2DES. At finite field, the simulation shows the density of the dot increasing monotonically as the bias is increased, but charge enters the dot more slowly when the system is in a Landau gap. When the bias is sufficiently large, incompressible bands of charge form at the edge, corresponding to filled Landau levels. At fixed bias, the total amount of charge in the dot oscillates as the magnetic field is swept. Electrons are alternatively added to the center of the dot and transferred out of the center into the edge regions. Once all the electrons have fallen into the lowest Landau level, the number of electrons in the dot decreases monotonically with any further increase in field strength.

\end{document}